\documentclass[conference]{IEEEtran}

\usepackage{stmaryrd}
\usepackage{graphicx}
\usepackage{color}
\usepackage{comment}
\usepackage{url}
\usepackage{wrapfig}
\usepackage{subfigure}
\usepackage{tabularx}
\usepackage{wrapfig}
\usepackage{balance}
\usepackage[justification=centering,small]{caption}

\usepackage{algpseudocode}
\usepackage{algorithm}
\usepackage{mathtools}

\newcommand{\superscript}[1]{\ensuremath{^{\textrm{#1}}}}

\def\wu{\superscript{*}}
\def\wg{\superscript{\dag}}

\begin{document}

\title{Adaptive 360 VR Video Streaming:\\ Divide and Conquer!}

\author{\IEEEauthorblockN{Mohammad Hosseini\wu\wg, Viswanathan Swaminathan\wg}
\IEEEauthorblockA{
%  \sharedaffiliation
  \begin{tabular}{ccc}
    \wu University of Illinois at Urbana-Champaign (UIUC) & &\wg Adobe Research, San Jose, USA \\
  \end{tabular}
  ~\\
Email: shossen2@illinois.edu, vishy@adobe.com}}

\maketitle

\begin{abstract}
While traditional multimedia applications such as games and videos are still popular, there has been a significant interest in the recent years towards new 3D media such as 3D immersion and Virtual Reality (VR) applications, especially 360 VR videos. 360 VR video is an immersive spherical video where the user can look around during playback. Unfortunately, 360 VR videos are extremely bandwidth intensive, and therefore are difficult to stream at acceptable quality levels. 

In this paper, we propose an adaptive bandwidth-efficient 360 VR video streaming system using a divide and conquer approach. We propose a dynamic view-aware adaptation technique to tackle the huge bandwidth demands of 360 VR video streaming. We spatially divide the videos into multiple tiles while encoding and packaging, use MPEG-DASH SRD to describe the spatial relationship of tiles in the 360-degree space, and prioritize the tiles in the Field of View (FoV). In order to describe such tiled representations, we extend MPEG-DASH SRD to the 3D space of 360 VR videos. We spatially partition the underlying 3D mesh, and construct an efficient 3D geometry mesh called \textit{hexaface sphere} to optimally represent a tiled 360 VR video in the 3D space. Our initial evaluation results report up to 72\% bandwidth savings on 360 VR video streaming with minor negative quality impacts compared to the baseline scenario when no adaptations is applied.

\end{abstract}
\IEEEpeerreviewmaketitle
%\vspace{-.5cm}

\section{Introduction}
\begin{wrapfigure}{R}{0.15\textwidth}
\setlength{\columnsep}{2pt}%
    \vspace{-.2cm}
  \centering
  \hspace{-.18cm}
  \includegraphics[width=.3\columnwidth]{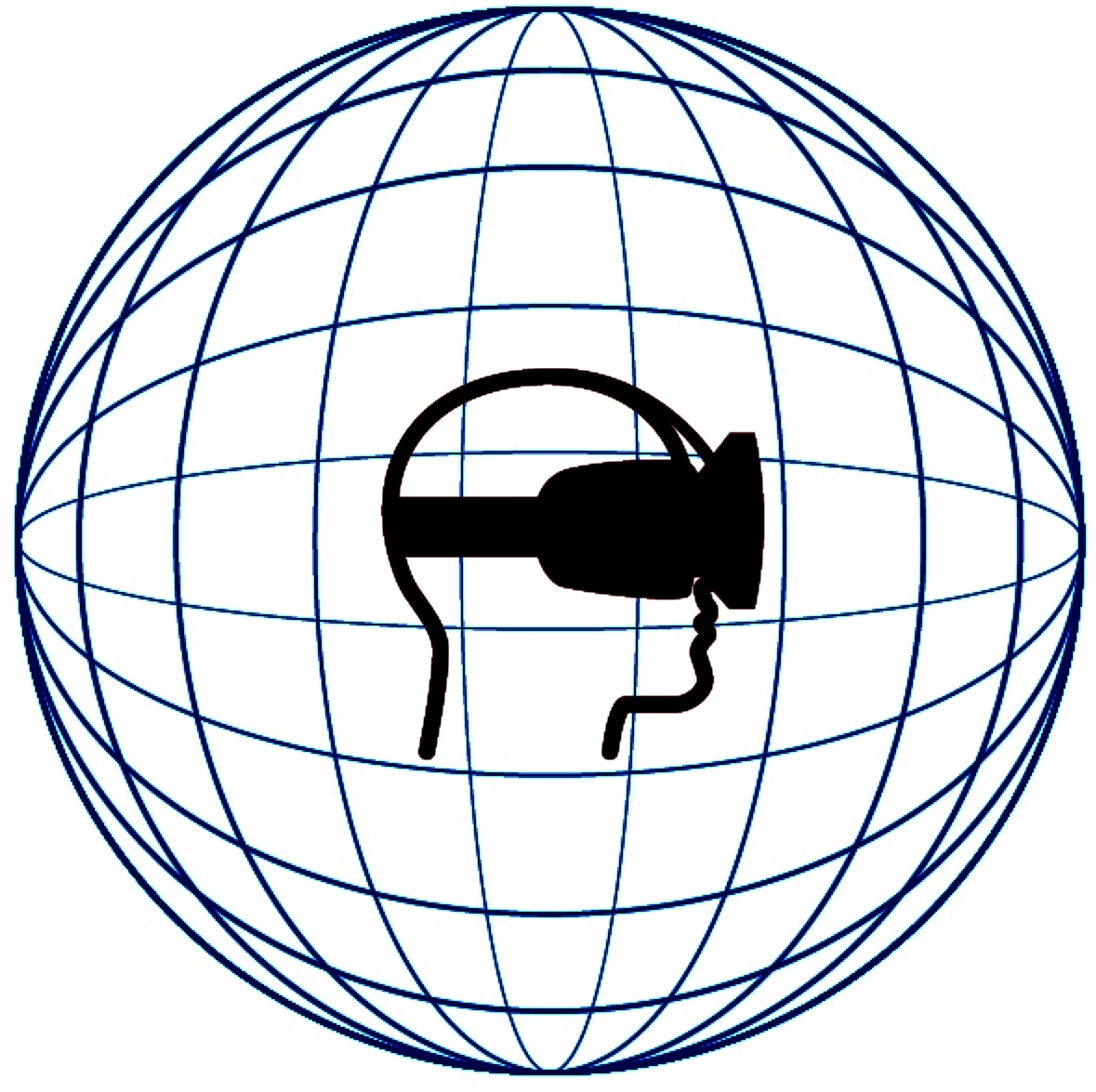}
    \caption{Visual structure of a 360 VR video} 
%\vspace{-.2cm}
\label{360video}
\end{wrapfigure}

Advances in computing hardware and networking technologies with support of high bandwidth have enabled the use of new 3D media such as 3D immersion and 360-degree VR video applications in the recent years. 360 VR videos are immersive spherical videos, mapped into a 3D geometry as shown in Figure \ref{360video}, where the user can look around during playback using a VR head-mounted display (HMD). This gives viewer a sense of depth in every direction.

Despite the promising nature of 360 VR videos, %many challenges still remain to deliver the VR content at high quality levels. While we are already witnessing the extreme growth of many high quality live video broadcasting services such as Youtube, 
existing 360 VR video applications are still restricted to lower resolutions compared to their 2D counterparts. Unfortunately, a major challenge is how to efficiently transmit the bulky 360 VR video streams to bandwidth-constrained devices such as wireless VR HMDs given their high bitrate requirements. Especially with the 4K video resolution being widely viewed as a functional minimum resolution for current HMDs, and 8K or higher being desired, these new media are extremely bandwidth intensive and difficult to stream at acceptable quality levels. Thus there must be a balance between the requirements of streaming and the available resources on the display devices. One of the challenges to achieving this balance is that we need to meet this requirement without much negative impact on the user's viewing experience. %One of the challenges to achieving this balance is that we need to meet the demands for a viewing preference given the limited bandwidth without much negative impact on the user's experience. 
While our work is motivated by the 360 VR video applications with 8K and 12K resolutions and the data rate issues that such rich multimedia system have, a semantic link between FoV, spatial partitioning, and stream prioritization has not been fully developed yet for the purpose of bandwidth management and high performance 360 VR video streaming. Hence, we propose to utilize this semantic link in our dynamic adaptations.
%\begin{figure}[!t]
%\centering
%\includegraphics[width=.4\columnwidth]{360video2.jpg}
%\caption{Visual structure of a 360 VR video}
%\vspace{-.5cm}
%\label{360video}
%\end{figure}

In this paper, we propose FoV-aware streaming adaptations for efficient delivery of high-resolution 360 VR videos to bandwidth-limited HMDs. We spatially divide the videos into multiple tiles while encoding and packaging, use MPEG-DASH SRD to describe the spatial relationship of tiles in the 360-degree space, and prioritize the tiles in the viewport. We then extend our tiling process to the 360-degree VR environment to \textit{conquer} the intense bandwidth requirements using viewport adaptation techniques. To achieve that, we spatially partition the underlying 3D mesh, and construct an efficient 3D geometry mesh called \textit{hexaface sphere} to optimally represent a tiled 360 VR video in the 3D space. On the other hand, due to the absence of a fine-grained prioritized mechanism, most 360 VR streaming systems today handle all the portions of the spherical view as equally important, resulting in sub-optimal resource usage. Our approach is to deliver higher bitrate content to regions where the user is currently looking and is most likely to look, and delivering lower quality level to the area outside of user's immediate viewport. Our initial evaluation results using a real-platform wireless HMD and multiple 360 VR video benchmarks show that our adaptations significantly reduces the amount of bandwidth required to deliver a high quality immersive experience, and increases the overall 360 VR video quality at a given bandwidth.

%The paper is organized as follows: in Section 2, we briefly cover background and some related work. Section 3 explains our methodology for 3D geometry construction and dynamic view-aware adaptation. Our experiments and evaluation results are presented in Section 4, while in Section 5 we conclude the paper and briefly discuss possible avenues for future work.

\section{Background and Related Work}
In this section we briefly present different concepts and categories of state-of-the-art related to our proposed approach.

\subsection{Dynamic Adaptive Streaming}
One of the main approaches for bandwidth saving on bandwidth-intensive multimedia applications is adaptive streaming. %Adaptive streaming is a process whereby the quality of a multimedia stream is altered in real time while it is being sent from a server to a client. This adaptation of quality is controlled by decision modules on either the client or the server. The adaptation may be the result of adjusting various network or device metrics. For example, with a decrease in network throughput, adaptation to a lower video bitrate may reduce video packet loss and improve the user's experience. 
%Researches on adaptive multimedia and HTTP video streaming ranges from network coding, rate determinations and quality of experience \cite{cmm1,cmm2,cmm3}. 
Dynamic Adaptive Streaming over HTTP (DASH) specifically, also known as MPEG-DASH \cite{dash1, dash} is an ISO standard that enables adaptive bitrate streaming whereby a client chooses the video segment with the appropriate quality (bit rate, resolution, etc.) based on the constrained resources such as bandwidth available at the client.%The multimedia content is stored on an HTTP server, and is accompanied by a Media Presentation Description (MPD) as a manifest of the available segments, their various bitrate alternatives, their URL addresses, and other characteristics.

As a part of ISO/IEC 23009-1:2015, a new ammendment called Spatial Relationship Description (SRD) has been added to MPEG-DASH standard, which allows to define spatial relationships between spatially partitioned multimedia content \cite{srd}. %The spatial parts can define a region of interest, or a tile, and can be represented by either an adaptation set or a sub-representation.
The SRD feature explores how to combine several spatially-related videos in a tiled combination, while at the same time provides backward compatibility with regular definition of adaptation sets. It provides a unique approach to tiled streaming in the perspective of the facilitated ultra high-resolution video display, specifically in the context of immersive environments such as those seen in 360 VR videos. There have been some work exploring the features of MPEG-DASH SRD. Le Feuvre \textit{et al.} in their work \cite{srd1} explored spatial access of 2D video contents within the context of MPEG-DASH, and discussed how the tiling feature of SRD extension can enable that. In another work \cite{srd2}, D'Acunto \textit{et al.} explored the use of MPEG-DASH SRD to partition a video into sub-parts to provide a zooming feature inside 2D video contents. In this work, we extend the semantics of MPEG-DASH SRD towards the context of 360 VR videos, and use that to partition the bulky 360 VR videos into spatially related tiles in the 3D space for the purpose of view-aware adaptation. Our aim is to explore prioritization of the tiles by assigning highest resolution only to those tiles that are relevant to the user's FoV and lower resolution to tiles outside of the user's FoV.

\subsection{Prioritized Multimedia Streaming}
Generally, different parts of multimedia can have different importance given various settings such as view, region, or the context. Hosseini \textit{et al.} \cite{hosseini1}, \cite{hosseini2} %, \cite{hosseini3} 
adopted prioritization techniques towards efficiently transmitting, rendering, and displaying bulky 3D contents to resource-limited devices given the importance of various 3D objects in the context. %DeVincenzi \textit{et al.} \cite{priority1} applied the concept of prioritization on a multi-user teleconference room equipped with a static camera capturing the whole room. When a speaker begins speaking, the system notifies the video encoder component to assign higher video resolution to the speaker's location in the scene. 
Similarly, in the context of 3D tele-immersive systems, the authors in \cite{mmsj} studied stream prioritization in regards to bandwidth savings. Their approach assigns higher quality to parts within users' viewport given the features of the human visual system.

In this paper, we build upon the concepts from these works to implement an adaptive prioritized view-aware streaming approach to reduce the bandwidth requirements of 360 VR video streaming. %To the best of our knowledge, there is currently no work that takes into account adaptive view-aware streaming of 360 VR videos using a prioritized tile-based approach.
\begin{figure}[!t]
\centering
\includegraphics[width=1.02\columnwidth]{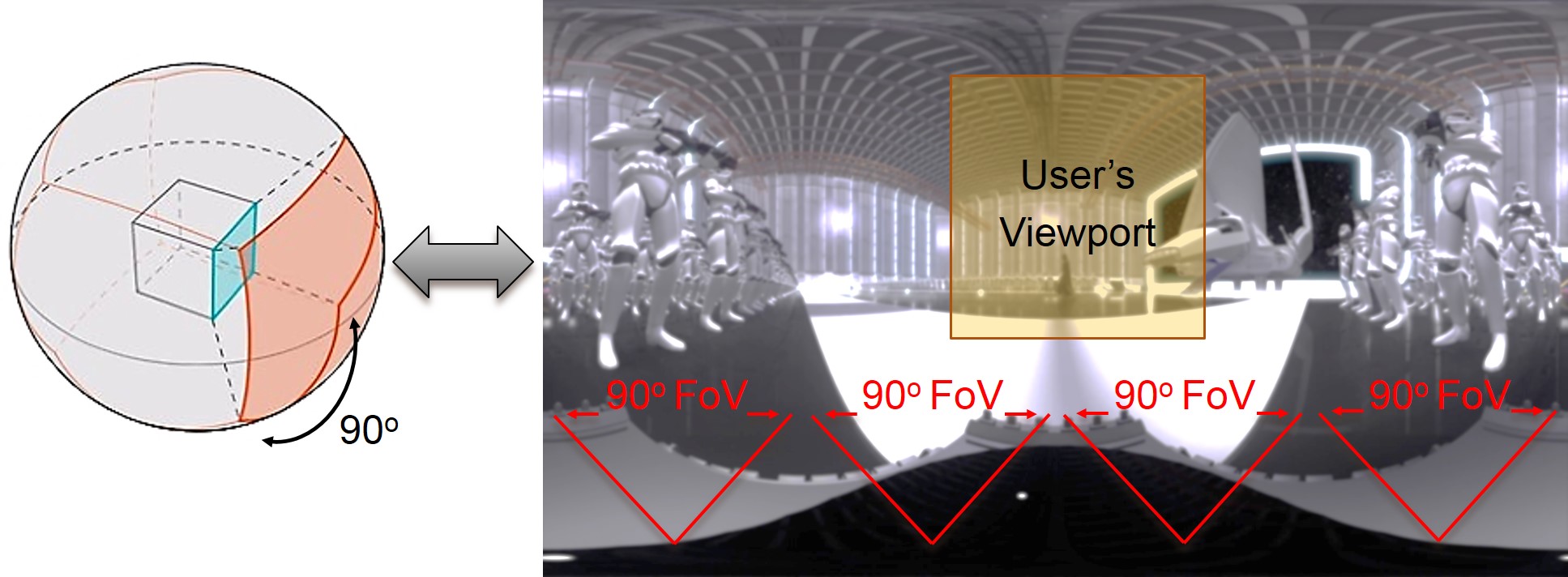}
\caption{An example visual view of how a 90-degree spherical viewport is mapped on a raw 360 video (\textit{Star Wars})}
\label{spherefov}
\vspace{-.5cm}
\end{figure}

\section{Methodology}
%\subsection{View-Aware Adaptation}
%https://dev.opera.com/articles/w3c-device-orientation-usage/
Similar to the context of 3D graphics, the visual experience of 360 VR videos is also based upon texturing. Textures are used to provide surface wrapping for a 3D object mesh, with 3D textures being a logical extension of the traditional 2D textures. 360 VR videos are created by mapping a raw 360 video as a 3D texture onto a 3D geometry mesh, often a sphere, with the user at the center of that geometry as shown in Figure \ref{360video}. In this 360-degree environment however, a user is viewing only a small portion of the whole raw 360-degree video at any given time. Therefore, a user's FoV is also only a small portion equivalent to a specific confined region on the 3D object mesh which is \textit{spatially} related to the corresponding portion of the raw content. For example, the Samsung Gear VR HMD offers a 96-degree FoV, meaning it can only almost cover a quarter of a whole 360-degree-wide content horizontally. Figure \ref{spherefov} illustrates this concept. The left subfigure shows an example 90-degree viewport as projected on a spherical 3D geometry, while the right subfigure shows how the mapping of the viewport corresponds to that of a given frame on a raw 360-degree video.

One of the major challenges in streaming 360-degree VR videos is the high bandwidth demands. %They are highly bandwidth intensive. 
To decrease the bandwidth requirements, our approach is to assign higher quality to parts within a user's viewport, and use lower quality to parts which are not within the immediate viewport of the user. This approach also makes it possible to stream tiles inside the viewport at highest resolution, at or near the native resolution of the HMD, virtually enabling a total resolution of 8K and higher in the 360-degree VR environment. To achieve that, our approach consists of two parts. First, the raw 360-degree video is spatially partitioned into multiple tiles. Using the features of MPEG-DASH SRD, a reference space is defined for each tile, corresponding to the rectangular region encompassing the entire raw 360-degree video. Second, we partition the underlying 3D geometry into multiple segments, each representing a subset of the original 3D mesh with a unique identifier. Finally, a mapping mechanism is defined for spatial positioning of the tiles on the 3D space, so that each tile be textured on its corresponding 3D mesh segment.

\subsection{3D Geometry Construction: Hexaface Sphere}

  \begin{wrapfigure}{R}{0.21\textwidth}
\setlength{\columnsep}{2pt}%
    \vspace{-.4cm}
  \centering
  \hspace{-.18cm}
  \includegraphics[width=.45\columnwidth]{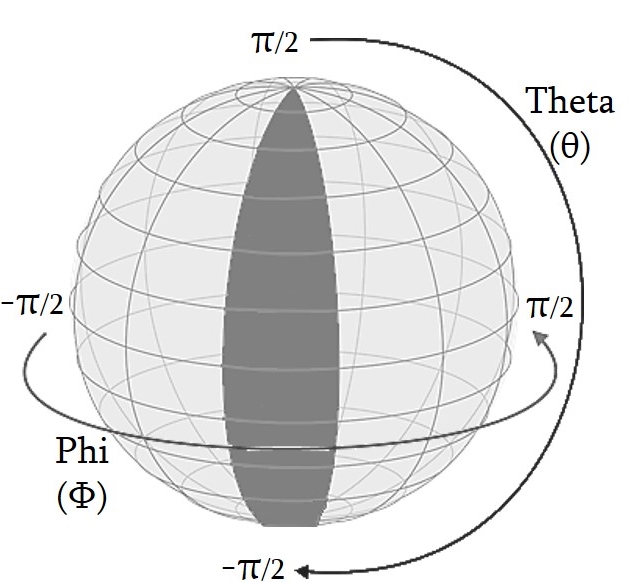}
\caption{A 3D sphere mesh. The highlighted area represents a slice.}
\label{sphere}
%\vspace{-.2cm}
\end{wrapfigure}
Most of the cameras available nowadays output 360 panoramic videos using a equirectangular format. While these videos are mapped into a flat projection for storage, they are inherently spherical. To achieve a spherical view as a common practice, a 3D sphere mesh is created surrounding the virtual camera inside the main virtual scene. Every individual photo-sphere, an image of a equirectangular 360 video frame, is wrapped on the internal surface of the sphere. The stereoscopic depth of 360 VR video requires two photo-spheres be stored and shown side-by-side representing the small disparity of left eye and right eye. The sphere is setup in such a way that it contains vertex locations and texture coordinates to achieve equirectangular mapping. The front faces of each of the rendered sphere polygons is culled to enable their internal surface showing as opposed to external showing. As a part of our geometry construction, we programmatically create a 3D sphere mesh in code to be able to control and further modify the geometry, and set the normal vectors of all of the mesh segments to point inside, towards the center of the sphere, achieving internal showing.

In order to create a 3D sphere mesh, we create an array of vertices by procedurally generating triangles for rendering. We approximate the sphere and provide a quality parameter to account for the trade-off between more smoothness and more triangles to render (representing highest quality) and rendering speed (lower spherical smoothness and chunkier shape). Figure \ref{sphere} illustrates a visual view of the structure of a 3D sphere mesh. Our quality parameter is a combination of two major parameters which together control the smoothness of the sphere: a) number of slices, which represent the number of vertical cuts. Let's assume each slice collides with a sphere's perimeter at a vertical degree $\theta$ which $-\pi/2\leq \theta \leq \pi/2$; and b) number of stacks, which is the number of horizontal cuts determining the number of rows of rectangles. Let's assume each stack collides with a sphere's perimeter at a horizontal degree $\phi$ which $-\pi/2\leq \phi \leq \pi/2$. Algorithm \ref{alg1} presents pseudo-code of our process to create 3D spherical mesh. %Figure \ref{missing-stack} clarifies the concepts of our 3D geometry construction by illustrating how a missing stack can affect the visual view of a 360 VR video on a real platform VR HMD.
%\begin{figure}[!t]
%\centering
%\includegraphics[width=.5\columnwidth]{thetaphi.jpg}
%\caption{A 3D sphere mesh. The highlighted area represents a slice.}
%\label{sphere}
%\vspace{-.3cm}
%\end{figure}

\begin{algorithm}[!t]
\begin{algorithmic}
\State $M$: the number of stacks
\State $N$: the number of slices
\State $\forall m : 0\leq m\leq M~and~\forall n: 0\leq n\leq N-1$, calculate \\and store a spatial point $P(x, y, z)$ such that:
\State $P_x \gets Sin(\pi \times m/M) \cdot Cos(2\pi \times n/N)$
\State $P_y \gets Sin(\pi \times m/M) \cdot Sin(2\pi \times n/N)$
\State $P_z \gets Cos(\pi \times m/M)$
\State Draw the line segments between the each vertex.
\end{algorithmic}
 \caption{Our process to generate a 3D sphere mesh}
 \label{alg1}
\end{algorithm}

%\begin{figure*}[!t]
%\centering
%\includegraphics[width=.8\textwidth]{missing-stack4.jpg}
%\caption{Example visual effect of a missing stack on our real-platform 360 video benchmark (Waldo).}
%\label{missing-stack}
%\vspace{-.3cm}
%\end{figure*}

%Given $M$ as the number of stacks and $N$ as the number slices, we calculate and store a spatial point $P(x, y, z)$ using the equations presented below:\\
%$\forall m : 0\leq m\leq M~and~\forall n: 0\leq n\leq N-1$,\\
%$P_x = Sin(\pi \times m/M) \cdot Cos(2\pi \times n/N)$,\\
%$P_y = Sin(\pi \times m/M) \cdot Sin(2\pi \times n/N)$,\\
%$P_z = Cos(\pi \times m/M)$,\\
%We store the points in a data structure and draw the line segments between the each vertex, accordingly. 
%Using too many stacks and slices exhausts hardware resources such as memory used for GPU-assisted video rendering, and therefore leads to lower performance without much improvement in quality.

Next, in order to generate 3D mesh segments, we partition the 3D sphere into multiple different 3D meshes, in a two-step process. In the first step, we split the sphere into 3 major parts:

\begin{itemize}
\item The top cap, which includes meshes from the top point of the sphere (i.e. top pole where $\theta=+\pi/2~^o$) to the top stack of the middle body ($\theta=+\beta^o$), totaling $(\pi/2 - \beta)~^o$,
\item The middle body which is a ring of triangles stretching from the bottom stack of the top cap ($\theta=+\beta^o$) to the top stack of the bottom cap ($\theta=-\beta ^o$), totaling $2\beta^o$,
\item The bottom cap, which includes triangles ranging from the bottom stack of the middle body ($\theta=-\beta^o$) to the bottom point of the sphere (i.e. bottom pole where $\theta=-\pi/2~^o$), for $\pi/2 - \beta$ degrees.
\end{itemize}

\begin{figure}[!t]
\centering
\vspace{-.3cm}
\includegraphics[width=.5\columnwidth]{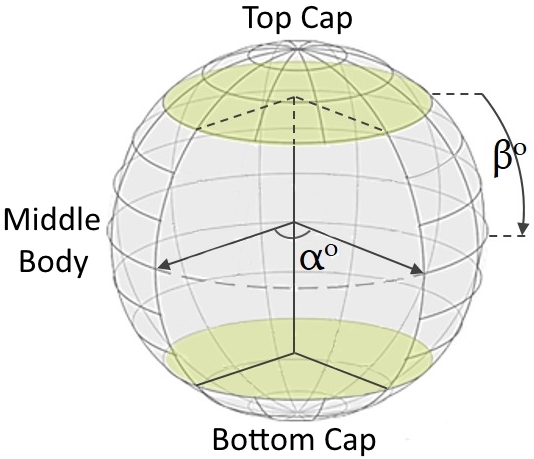}~~\includegraphics[width=.4\columnwidth]{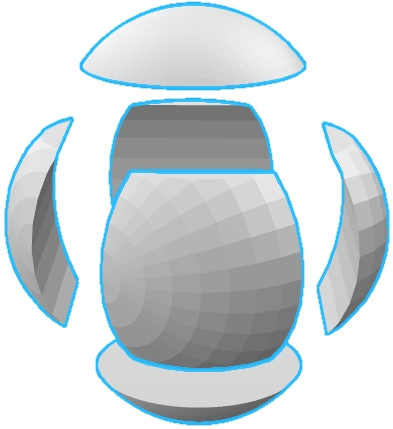}
\caption{Visual overview of a generated hexaface sphere.}
\label{hexaface}
\vspace{-.5cm}
\end{figure}

360-degree videos formatted in equirectangular projection can contain redundant information at the top and bottom ends, but not the middle body. So in the second step, we further split the middle body into multiple 3D meshes, each covering $\alpha^o$ of the entire 360-degree wide screen. The number of second-step partitions, and therefore $\alpha^o$, can be a function of the horizontal FoV of the VR HMD platform. %Taking the Samsung Gear VR HMD as our experimental VR device, it can cover almost a quarter of an entire 360-degree-wide content horizontally.
We split the middle body into four segments with $\alpha^o=\pi/2$, so that each segment has a 90-degree cover, and similarly set $2\beta=\pi/2$ following the default settings for the vertical FoV. Overall, with this procedure, our projection will result into a combination of \textit{six} 3D meshes of a 3D sphere. We call our customized 3D geometry a \textit{hexaface sphere} 3D mesh. Figure \ref{hexaface} illustrates a visual overview of our process to generate a hexaface sphere. It should be noted that our assignment of $\alpha=\pi/2$ and $\beta=\pi/4$ is not a hard requirement, and is derived from the default settings of our VR HMD device as well as the result of our performance and quality trade-offs. Our approach is general, and depending on the performance of underlying hardware can be extended towards higher number of tiles and 3D meshes. Similarly, our tiling process is general, and therefore can also be employed in other geometrical projections such as cube maps.
 
\subsection{Viewport Tracking}
VR device orientation is usually defined using three types of rotation over the $X$, $Y$, and $Z$ axes. The values are inherently represented in the \textit{Tait-Bryan} anglular system, which uses a special form of Euler angles that require 3 rotations around each of the 3 axes. %While using Euler angles for orientation tracking is a good way to visualize the rotations of a VR device as it moves around in physical space, it introduces an issue known as \textit{gimbal lock} when modeling 3-axis rotations \cite{gimbal}. As two of the circular rotation planes (i.e. gimbal) start approaching the same parallel direction, there will be no determination of which way to turn. When this occurs, one degree of freedom is effectively lost for the 3-axis modeling process of device orientation.
To avoid possible orientation tracking problems such as \textit{gimbal lock}, the angular Euler system is transformed into a different rotational system known as a \textit{quaternion}, which is converted into a unit rotation matrix. %A quaternion consists of two sets of values: Firstly, a set consisting of a $[x, y, z]$ tuple representing the axis about which the device rotation occurs. And secondly, a $w$ component representing the amount of rotation that will occur about this axis. With these four values it is possible to describe device orientation perfectly while also avoiding introducing the gimbal lock problems. Overall, 
To enable view awareness, we use the following three steps to create valid confines of unit quaternions specifically set for each of the hexaface sphere 3D mesh segments:
\begin{itemize}
\item convert Euler angles to a unit quaternion representation for VR device orientation tracking,
\item calculate an array corresponding to a normalized direction vector from our quaternion,
\item combine the values together to compute the confines of segment-specific quaternion representations inside the hexaface sphere.
\end{itemize}

With the confines of each 3D mesh segment being defined, we then identify which segments and the corresponding video tiles intersect with a user's viewport and implement our viewport tracking at every frame. With viewport tracking, we then implement view-aware adaptation, and dynamically deliver higher bitrate content to the tiles within the user's FoV, and assign lower quality level to the area outside the user's immediate viewport.

\section{Evaluation}
%\begin{figure}[!t]
%\centering
%\includegraphics[width=.4\textwidth]{karate-top.jpg}\\
%~\\
%\includegraphics[width=.1\textwidth]{karate-middle1.jpg}~~~\includegraphics[width=.1\textwidth]{karate-middle2.jpg}~~~\includegraphics[width=.1\textwidth]{karate-middle3.jpg}~~~\includegraphics[width=.1\textwidth]{karate-middle4.jpg}\\
%~\\
%\includegraphics[width=.4\textwidth]{karate-bottom.jpg}\\
%\caption{Various tiles of an example 360 video (Karate) according to the six 3D meshes of our hexaface sphere geometry.}
%\label{karate}
%\vspace{-.3cm}
%\end{figure}
To evaluate our work, we used Samsung Gear VR HMD mounted with the Samsung Galaxy S7 smartphone with 4GB RAM and Android Marshmallow 6.0.1 as our target VR platform. We used Oculus Mobile SDK 1.0.3 along with Android SDK API 24 for development of a 360 VR video streaming application prototype based on MPEG-DASH SRD. Our VR platform provides a total resolution of 2560x1440 (1280x1440 per eye), with maximum frame rate of 60 FPS and a horizontal FoV of 96 degrees. As stated before, we set the vertical FoV of our 360 VR video prototype to 90 degrees. We prepared 5 different 360 equirectangular-formatted sample videos publicly available on Youtube as test sequences for the purpose of applying our adaptations. Table \ref{videos} provides detailed information about our test video sequences.

  \begin{wrapfigure}{R}{0.2\textwidth}
\setlength{\columnsep}{5pt}
    \vspace{-.2cm}
  \centering
  \hspace{-.2cm}
  \includegraphics[width=.45\columnwidth]{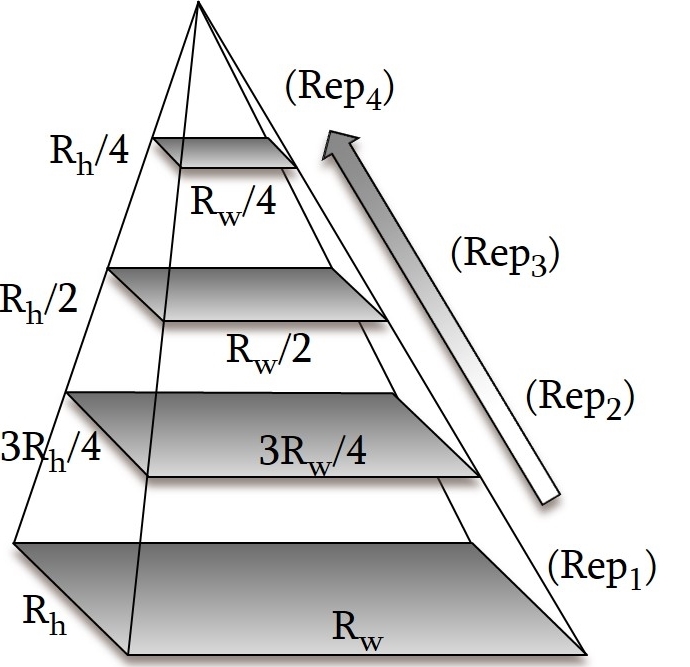}
\caption{Our hierarchical resolution degrading for various representations.}
\label{hierarchy}
%\vspace{-.2cm}
\end{wrapfigure}

\begin{table}[!t]
\vspace{-.3cm}
\label{videos}
\caption{}
\centering
\begin{tabular}{llll}
\hline
360 Video    & Resolution & Orig. Bitrate (Mbps) & FPS (Hz) \\
\hline
Waldo      & 3840x1920     & 20.0 & 30      \\
Plane          & 3840x1920     & 13.1 & 30         \\
Karate       & 1920x960     & 4.4    & 30  \\
Star Wars       & 3840x2160     & 13.0 & 24     \\
Racing Car  & 4096x2048     & 12.6 & 24   \\
\hline
\end{tabular}
\vspace{-.3cm}
\end{table}
To benefit from the features of MPEG-DASH SRD-based streaming and for the purpose of texturing the hexaface sphere mesh, we developed an executable script based on FFmpeg Zeranoe 64-bit API that given a video input, spatially crops the video and generates 6 different tiles as per requirements of our hexaface sphere geometry. We encoded all video segments with H.264 AVC encoder into 4 different representations using a hierarchical resolution degrading, with original resolutions decreasing polynomially at every step, as depicted in Figure \ref{hierarchy}. % (it should be noted that resolution degrading is not a hard requirement. We could have degraded just the quality and the overall bitrate instead of changing resolutions). Assuming $R_w(\tau_{REP_i})$ be the resolution width (and similarly resolution height $R_h(\tau_{REP_i})$) of a specific video tile $\tau$ for the representation $i$, then:
%\begin{equation}
%R(\tau_{REP_{i}}) = R(\tau_{REP_{1}}) \cdot \frac{(4-(i-1))}{4} ~~~ \forall i : 1\leq i\leq L.
%\end{equation}

%We used a Dell XPS 8900 x64-based PC mounted with Microsoft Windows 7 Professional version 6.1.7601 OS as the HTTP-based streaming server, and used MPEG-DASH SRD to describe our tiling.
%In this initial study, we used the Samsung Galaxy S7 smartphone as the local streaming server to filter the negative impacts of network latency and bandwidth variations on the experimental results, and purely focus on the amount of bandwidth savings. 
%Figure \ref{karate} illustrates how our tiling process is applied against an example frame within the \textit{Karate} 360 video sequence, according to the 3D meshes of our hexaface sphere geometry as shown in Figure \ref{hexaface}.

%\begin{figure}[!t]
%\centering
%\includegraphics[width=.6\columnwidth]{hierarchy2.jpg}
%\caption{Our hierarchical resolution degrading for various representations.}
%\label{hierarchy}
%\vspace{-.3cm}
%\end{figure}

\begin{figure}[!t]
\centering
\includegraphics[width=\columnwidth, trim = 50 260 50 250, clip = true]{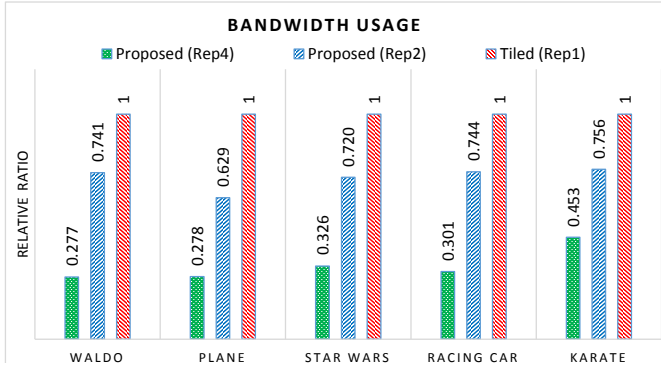}
\caption{A comparison of bandwidth savings of streaming different 360 VR videos, using our adaptations, and tiled-streaming with no adaptation.}
\label{results}
\vspace{-.5cm}
\end{figure}

\begin{figure}[!t]
\centering
\includegraphics[width=.5\textwidth]{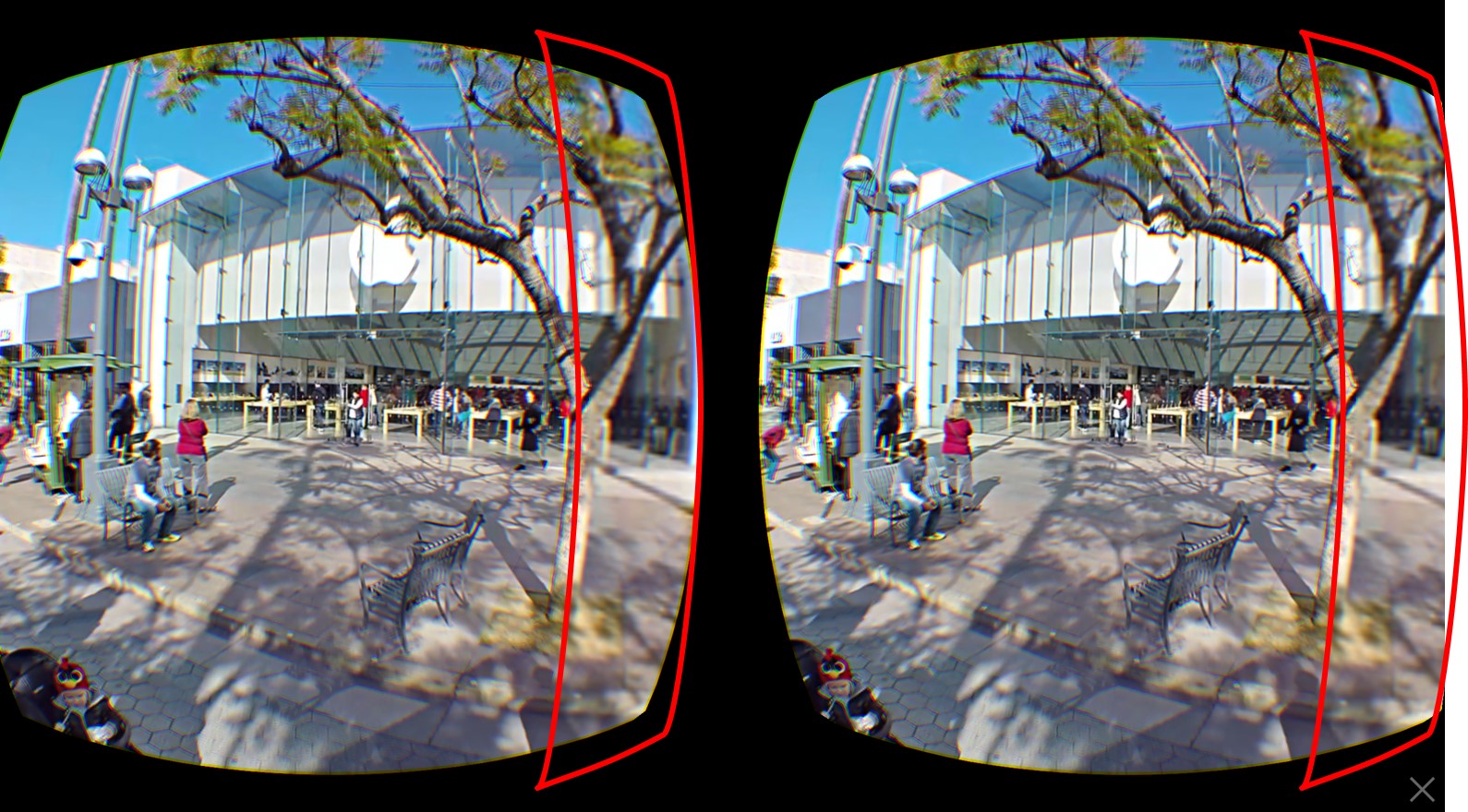}\\
\caption{Visual comparison of a specific frame within \textit{Waldo} with the peripheral tiles having $REP_4$ (resolution of 240x480)} %(Bottom) $REP_2$ with resolution of 720x1440
\label{240}
\vspace{-.3cm}
\end{figure}

%For the purpose of quantitative analysis,%quality measurement, we assume that the quality of a 360 VR video session is a function of bitrates for all tiles (with maximum quality corresponding to maximum resolutions of all delivered tiles). Our approach is general and is independent from the measure of quality. In this initial study, we use the simplest of these functions- the average bitrate. 
We applied various sets of resolutions to different tiles to experiment how our prioritized adaptations affects the average bitrate as well as the perceived quality. %We collected statistics for the average bitrate, and compared our results with the current practices in 360 VR video players where no tiled-based approach is employed. 
Each trial of our experiment was run for a total of 30 seconds, and during each run we setup the experiment in such a way that each tile is within user's viewport for 5 seconds. % We repeated each experiment several times to ensure that the standard deviations are within acceptable limits. 
We measured the bandwidth usage in terms of average bitrate, when maximum resolution is assigned for tiles within immediate FoV, and lower resolutions assigned to the other tiles. Figure \ref{results} demonstrates results for only a small subset of our experiments on all of our benchmarks, with ratios normalized to fit within a unit. We measured the relative bandwidth usage when using our adaptations compared to the baseline case where no adaptation is applied (the 360 VR video is tiled; no viewport awareness is present, and all tiles are streamed with highest representation ($REP_1$). As can be seen, the results show that our adaptations can significantly save bandwidth usage for up to 72\% compared to baseline case where our adaptation approach is not employed. Figure \ref{240} shows a sample screenshot of the experiments on \textit{Waldo}. While the highest representation possible ($REP_1$- resolution of 960x1920) is delivered to the main front tile, lowest representation is delivered to the peripheral tile on the right identified by the red confines ($REP_4$- resolution of 240x480) %, whereas in Figure \ref{240} (bottom), the peripheral tile on the right has second highest representation assigned to it ($REP_2$- resolution of 720x1440). The red confines specify the approximate area for the peripheral tile with the lower quality. 
Our adaptations results in minor noticeable quality impacts, sometimes not even perceptible, while maintaining the original quality for the main viewport to ensure a satisfactory user experience.
%As can be seen in the screenshot (and have been witnessed during our experiments), even the lowest representation on the peripheral tiles not within immediate viewports results in minor visual changes from a user's perspective, sometimes not even perceptible, yet at the same time it maintains the original quality for the main viewport to ensure a satisfactory user experience. The tiling feature of our approach can further increase the overall quality of 360 VR videos, virtually enabling a total 360-degree video resolution of 8K and higher in the 360-degree space. %It makes it possible to stream tiles within the viewport at even higher resolutions, at or near the native resolution of the HMD, which previously was not possible due to the limited hardware resources processing a single bulky video content.
%Overall, considering the significant bandwidth saving achieved using our adaptations, it is reasonable to believe that many 360 VR video users would accept such minor visual changes given their limited bandwidth. The results of our user study will be presented in details in the future work.

\section{Conclusion and Future Work}
%360 VR video provides an immersive 360-degree video experience where the user can look around during playback. Unfortunately, due to the tremendous bandwidth bandwidth requirements, these new media are difficult to stream at acceptable quality levels.
In this paper, we proposed bandwidth-efficient FoV-aware streaming adaptations to tackle the high bandwidth demands of 360 VR videos. Our novel adaptations exploits the semantic link of MPEG-DASH SRD with a user's viewport to provide dynamic view awareness in the context of VR videos. We \textit{divide} the bulky 360 VR videos as well as the underlying 3D geometry into spatially partitioned segments in the 3D space, and then \textit{conquer} the huge streaming bandwidth requirements using a dynamic viewport adaptation. Our initial experimental results shows our adaptations can save up to 72\% of bandwidth on 360 VR video streaming without much noticeable quality impacts.

Appendix I includes our adaptive rate allocation algorithm for tiled streaming given the available bandwidth.

In the future, we also plan to extend the hexaface sphere towards higher number of 3D segments and measure the bandwidth saving-performance trade-offs. 

\bibliographystyle{IEEEtran}
\bibliography{ref}

\newpage
\section{Appendix I: Rate Allocation Problem}
Our proposed view-aware tile-based 360 VR video streaming system is ``adaptive'' in the sense that it priorities the tiles. It delivers higher bitrate content to the tiles where the user is currently looking at and is most likely to look at, and lower quality level to the tiles outside of a user's immediate viewport. For that purpose, we hereby explain the rate allocation problem within the context of tiled 360 VR video streaming, and propose a rate allocation heuristic algorithm to enable the adaptations within our tile-based 360 VR video streaming system. This appendix aims to mathematically model, and propose a rate allocation heuristic algorithm within the context of tile-based 360 VR video streaming to VR devices.

The rate selection and allocation problem is the well-known binary Knapsack optimization problem, for which one approach to tackle is to transmit a subset of the whole tiles within the 360-degree VR environment. The binary Knapsack problem is NP-hard, but efficient approximation algorithms can be utilized (fully polynomial approximation schemes), so this approach is computationally feasible. However, using this method only a \textit{subset} of the whole tiles are selected, which is not desired since the VR user intends to receive \textit{all} the necessary tiles to avoid black views when the user turns head and changes orientation in the 360-degree space. Our proposed algorithms selects all necessary tiles, but with different bitrates according to their priorities given the user's viewport. This is a \textit{multiple-choice knapsack problem (MCKP)} in which the items (tiles in our context) are organized into groups corresponding to the objects. Each group contains the highest bitrate stream corresponding to an object and lower-bitrate versions of the same stream given the adaptation manifest. 

\begin{figure}[!t]
\centering
\includegraphics[width=.5\columnwidth]{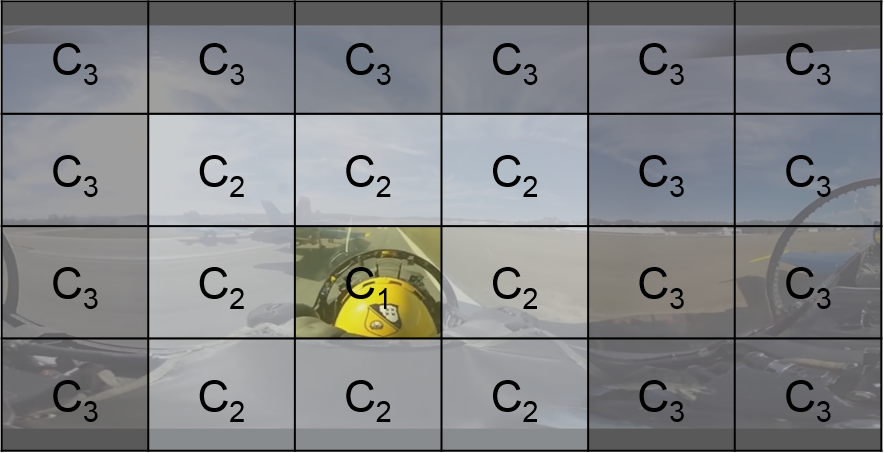}
\caption{An example tile prioritization. Tiles within viewport are assigned highest priority ($C_1$), while tiles outside of viewport are assigned lowest priority ($C_3$)}
\label{fig:priority}
\end{figure}

There are $n$ tiles $\mathcal{T}=\{\tau_1,\tau_2,\ldots,\tau_n\}$ in the 360-degree VR environment. The highest possible representation of each $\tau_h \in \mathcal{T}$ has a bitrate requirement of $s_{\tau_h}$, and a priority or importance coefficient $p_{\tau_h}$ given the various priority classes per our definition. With view awareness feature of our tiling process, our algorithm assigns highest priority to the tiles within the user's immediate viewport ($C_1$), and lowest priority ($C_3$) to the tiles in the area outside the user's viewport in the 360-degree space. Figure \ref{fig:priority} illustrates how our prioritization approach is applied against tiles in the context of tiled 360 degree VR videos. Our approach is general, and can work with any number of priority classes. We use three classes in this pilot study. We assume the quality contribution of a tile $\tau_h$ is a simple function $q_{\tau_h} = p_{\tau_h}\times s_{\tau_h}$. The available bandwidth in every interval limits the total bitrate of all tiles that can be received at the headset device to $W$, which serves as an available budget.

Let $X= \{x_1,x_2,\ldots,x_n\}$, be the set of tiles that are received at the headset device, serving as the output of running rate allocation algorithm. Each $x_i \in X$ corresponds to an original tile $\tau_i \in \mathcal{T}$. Similarly, each $x_i$ has a priority coefficient $p_{x_i} = p_{\tau_i}$.

We assume there are $L$ number of representations avialable given the manifest, with a representation of level $k$ noted as $R_k$ ($0 \leq k \leq L$) and the bitrate of a tile $\tau_i$ with representation $R_k$ noted as $s_{\tau_i}^{R_k}$. We assume the lowest bitrate corresponds to the representation with highest ID which is $R_L$ which is determined as the \textit{minimum bitrate} that can be tolerated by users. In a similar way, the quality contribution of a tile $x_i$ is $q_{x_i}=p_{x_i}\times s_{x_i}$.

\subsection{Heuristic Algorithm}

\begin{algorithm*}[!t]
\begin{algorithmic}
\State $\mathcal{T}$: prioritized list of tiles sorted from highest to lowest priority
\State $\tau_i$: tile with highest bitrate $s_{\tau_i}$
\State $x_i$: adapted tile with bitrate $s_{x_i}$
\State $L$: Number of representation levels
\State $R_L$: Level $L$ representation
\State Calculate $W_{min} = \sum s_{\tau_i}^{R_L}$ \%comment: minimum bitrate requirement for all tiles
\State $\forall \tau_i \in \mathcal{T}: s_{x_i} \gets s_{\tau_i}^{R_L}$ \%comment: assign $R_L$ (minimum bitrate) to all $\tau_i$'s.
\State $W_0 \gets W - W_{min}$ \%comment: initialization
\While {$s_{\tau_i}- s_{\tau_i}^{R_L} \leq W_{i-1}$}
\%comment: i=1 initially.
\State $s_{x_i} \gets s_{\tau_i}$
\State $W_i \gets W_{i-1} - (s_{\tau_i}- s_{\tau_i}^{R_L})$
\State{ $i \gets i+1$~~\%comment: adapt next tile}

\EndWhile \\
\%comment: above loop repeats until a tile $\tau_{\ell}$ cannot be delivered at highest bitrate within the remaining bandwidth budget $W_{\ell-1}$.
\State $\ell \gets i$ \%comment: resulting from above loop.
\State{Find lowest $L' \leq L$ such that \\$s_{\tau_\ell}^{R_{L'}} \leq W_{\ell-1}+ s_{\tau_\ell}^{R_L}$ ~~\%comment: determines the highest bitrate possible at which $\tau_{\ell}$ can be received within remaining budget, by calculating the lowest representation level $L'$.}
\State{ $s_{x_{\ell}} \gets s_{\tau_\ell}^{R_{L'}}$ \%comment: adapt $\tau_{\ell}$ and calculate $s_{x_{\ell}}$}
\end{algorithmic}
 \caption*{Algorithm 2: Rate allocation heuristic algorithm}
 \label{algorithm2}
\end{algorithm*}

Let $S$ be the total bitrate of all streams, and $W$ be the available bandwidth budget. The minimum quality that the user can tolerate is given as the representation of level $L$ noted as $R_{L}$. Let $C_{1}$, $C_{2}$, and $C_{3}$ be the class of tiles with the highest priority, medium priority, and lowest priority, respectively.

For each tile $\tau_i$ in $\mathcal{T}$, we calculate $q_i$ as described previously. This is the contribution that $\tau_i$ would make to the average quality of the system if it were received at highest bitrate possible. We then calculate $W_{\min} = \sum s_{\tau_i}^{R_L}$ which is the minimum bitrate that is needed to receive all tiles at their lowest bitrates. In the following, assume that $W_{\min} \leq W$ so the unused bitrate budget would be $W_0=W-W_{\min}$.

To determine the best bitrate for each tile, our algorithm sorts the prioritized list of tiles by the global priority from the largest to the smallest. For ease of notation in the following, suppose that the tiles are re-indexed so that the sorted list of tiles is $\tau_1,\tau_2,\ldots,\tau_n$. If $s_{\tau_1}- s_{\tau_1}^{R_L}\leq W_0$ then there is enough unused budget to receive $\tau_1$ at highest bitrate ($R_0$), so the tile $x_1$ would have $s_{x_1}=s_{\tau_1}$ and would contribute $q_1$ to the average quality. This leaves an unused bandwidth budget of $W_1=W_0 - s_{\tau_1}^{R_0}- s_{\tau_1}^{R_L}$ for the remaining tiles after $x_1$. The algorithm repeats for $\tau_2, \tau_3,\ldots$ until some tiles $\tau_{\ell}$ cannot be received at highest bitrate within the remaining budget $W_{\ell-1}$. It then determines the highest possible bitrate at which it can be received by calculating the lowest representation level $L': L'\leq L$ such that $s_{\tau_\ell}^{R_{L'}} \leq W_{\ell-1}+ s_{\tau_\ell}^{R_{L}}$. The tile $x_{\ell}$ will have bitrate $s_{x_{\ell}} = s_{\tau_\ell}^{R_{L'}}$ and will contribute $q'_{\ell}$ to the average quality of the whole. The remaining bandwidth budget after streaming $x_{\ell}$ will be $W_{\ell} = W_{\ell-1} - s_{\tau_\ell}^{R_{L'}}$. The algorithm repeats this process to determine the proper bitrates, amount of bandwidth budget, and quality contribution for each of the remaining tiles $x_{\ell+1},x_{\ell+2},\ldots,x_{n}$. Algorithm \ref{algorithm2} describes our heuristic algorithm.

\begin{figure}[!t]
\centering
\includegraphics[width=\columnwidth]{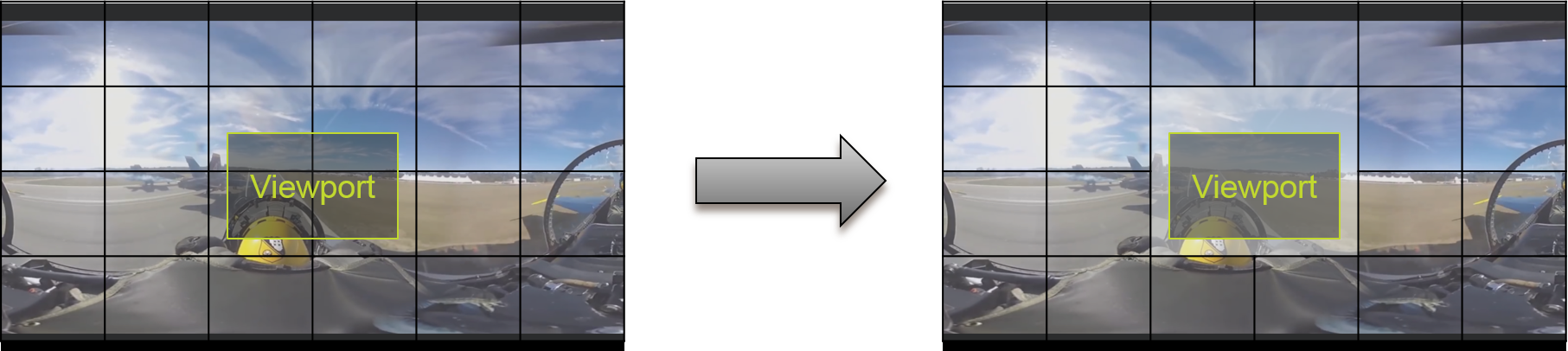}
\caption{Multi-tile adaptation within an example 360 video frame. Multiple tiles within viewport are assigned same priority.}
\label{fig:multi}
\end{figure}

As a special feature, our algorithm considers cases where the viewport intersects with multiple tiles in the 360-degree VR environment. Figure \ref{fig:multi} illustrates the concept. As a differentiating tweak in that scenario, our multi-tile rate allocation heuristic ensures all visible tiles (tiles within a user's viewport) get the same bitrate to avoid edging problems due to quality variation of tiles.  For such cases, our algorithm packs all tiles within the viewport together and deals with them as a single tile just for the purpose of rate allocation, so that all visible tiles are assigned the same bitrate. This tweak minimizes any possible edged views on the tile boundaries, therefore enabling a smooth viewing experience in the tiled 360 VR video scenario. Treating all visible tiles as a single tile depends on the developer preference and is not a hard requirement. It can be a pack of any of visible tiles given how much of the visible tiles intersect with the viewport.

The algorithm needs a one-time implementation in the beginning of the session for the main process. Therefore it is implemented in real-time and does not provide any additional overhead during the runtime. It is implemented efficiently in $O(n log n)$ time and $O(n)$ space and produces solutions close optimal. The approximation error depends on the difference between the bitrate chosen for the first tile that cannot be received at highest bitrate (i.e. $\tau_\ell$) and the remaining budget available to receive it.
\end{document}